\documentclass[namedreferences]{SolarPhysics}
\usepackage[optionalrh]{spr-sola-addons} 
\usepackage{graphicx}        
\usepackage{color}           
\usepackage{url}             

\usepackage{longtable}

\newcommand{\grad}{ {\bf \nabla } }

\newcommand{\aap}{    {\it Astron. Astrophys.}}

\newcommand{\apj}{    {\it Astrophys. J.}}

\newcommand{\jgr}{    {\it J. Geophys. Res.}}

\newcommand{\solphys}{{\it Solar Phys.}}

\begin{document}

\begin{article}

\begin{opening}

\title{Multiscale Edge Detection in the Corona}

\author{C.~Alex~\surname{Young}$^{1}$\sep
        Peter~T.~\surname{Gallagher}$^{2}$
       }
       
\runningauthor{C.A. Young, P.T.  Gallagher}
\runningtitle{Multiscale Edge Detection in the Corona}

   \institute{$^{1}$ ADNET Systems Inc., NASA/GSFC, Greenbelt, MD, 20771, USA\\
                     email: \url{c.alex.young@nasa.gov}\\ 
              $^{2}$ Astrophysics Research Group, School of Physics, Trinity College Dublin, Dublin 2, Ireland}

\date{Received: 7 December 2007 / Accepted: 17 March 2008}              

\begin{abstract}

Coronal Mass Ejections (\,CMEs\,) are challenging objects to detect using automated
techniques, due to their high velocity and diffuse, irregular morphology. A
necessary step to automating the detection process is to first remove the
subjectivity introduced by the observer used in the current, standard, CME
detection and tracking method. Here we describe and demonstrate a multiscale
edge detection technique that addresses this step and could serve as one part
of an automated CME detection system. This method provides a way to objectively
define a CME front with associated error estimates. These fronts can then be
used to extract CME morphology and kinematics. We apply this technique to a CME
observed on 18 April 2000 by the Large Angle Solar COronagraph experiment (\,LASCO\,) C2/C3 and a CME
observed  on 21 April 2002 by LASCO C2/C3 and the  {\it{Transition Region and Coronal Explorer}}
(\,TRACE\,). For the two examples in this work, the heights determined by the standard manual
method are larger than those determined with the multiscale method by  $\approx{10}\%$ using LASCO data and
$\approx{20}\%$ using TRACE data.
\end{abstract}
\end{opening}

\section{Introduction}
     \label{S-Introduction}

Currently, the standard method for detection and tracking of Coronal Mass Ejections (\,CMEs\,) is by visual
inspection ({\it e.g.}, the CUA CDAW CME catalog, available at \url{http://cdaw.gsfc.nasa.gov}). A human
operator uses a sequence of images to visually locate a CME. A feature of
interest is marked interactively so that it may be tracked in the sequence of
images. From these manual measurements, the observer can then plot
height--time profiles of the CMEs. These height measurements are used to
compute velocities and accelerations ({\it e.g.}, \opencite{gallagher03}).  Although the
human visual system is very sensitive, there are many problems with this
technique. This methodology is inherently subjective and prone to error. Both
the detection and tracking of a CME is dependent upon the experience, skill, and
well being of the operator. Which object, or part of the CME
to track varies, from observer to observer and is highly dependent on the quality
of the data. Also, there is no way to obtain statistical uncertainties, which is
particularly important for the determination of velocity and acceleration
profiles. Lastly, the inability to handle large data volumes and the use of an
interactive, manual analysis do not allow for a real-time data analysis
required for space weather forecasting. A visual analysis of coronagraph data is a
tedious and labor-intensive task. Current data rates from the {\it Solar and Heliospheric
Observatory} (\,SOHO\,) are low enough to make an interactive analysis possible (${<}$1
Gigabyte/day). This will not be the case for recent missions such as the {\it Solar
TErrestrial RElations Observatory} (\,STEREO\,) and new missions such as the {\it Solar
Dynamics Observatory} (\,SDO\,). These missions have projected data rates that make
an interactive analysis infeasible ($>$1 Terabyte/day). For these reasons, it is
necessary to develop an automatic, real-time CME detection and tracking system.
Current digital imaging processing methods and paradigms provide the tools
needed for such a system. In the next section we discuss in general, the parts needed in
an automated system, as well as some of the current work on this topic.

\section{A System for Automatic CME Detection} 
      \label{S-autodetect}      

The general design for an automated CME detection system should basically follow
the digital image-processing paradigm described by \inlinecite{gonzalez02}.
Digital image processing as they describe can basically be broken into three
parts: {\it{i}}) image preprocessing and segmentation; {\it{ii}}) image representation and
description, and; {\it{iii}}) object recognition. Image preprocessing includes standard
image preparation such as calibration, cosmic ray removal but it also includes
noise reduction based on the statistics of the image ({\it e.g.}, Gaussian or Poisson;
\inlinecite{starck02}). Image segmentation is the extraction of individual
features of interest in an image ({\it e.g.}, edges, boundaries, and regions). Some
methods used for this part include filtering, edge detection, and morphological
operations.  Image representation and description converts the extracted
features into a form such as statistical moments or topological descriptors
({\it e.g.}, areas, lengths, {\it{etc.}}) that are easier to store and manipulate
computationally. Object recognition includes techniques such as neural networks
and support vector machines (\,SVMs\,) to characterize and classify descriptors
determined in the previous step.

Determining the complex structure of CMEs is complicated by the fact that CMEs
are diffuse objects with ill-defined boundaries, making their automatic
detection with many traditional image processing techniques a difficult task. To 
address this difficulty, new image processing methods were employed by
\inlinecite{stenborg03} and \inlinecite{portier01}, who were the first to apply a
wavelet-based technique to study the multiscale nature of coronal structures in
LASCO and EIT data, respectively. Their methods employed a multilevel
decomposition scheme via the so-called ``{\it \`a trous}'' wavelet transform.

\inlinecite{robbrecht04},  developed a system to autonomously detect CMEs in image
sequences from LASCO. Their software, {{Computer Aided CME Tracking}}
(\,CACTus\,)(\url{http://sidc.oma.be/cactus/}), relies on the detection of bright ridges in
CME height--time maps using the Hough transform. The main limitation of this
method is that the Hough transform (as implemented) imposes a linear height--time
evolution, therefore forcing constant velocity profiles for each bright feature.
This method is therefore not appropriate to study CME acceleration. Other
autonomous CME detection systems include ARTEMIS \cite{boursier05}
(\url{http://lascor.oamp.fr/lasco/index.jsp}), the {Solar Eruptive Event
Detection Systems} (\,SEEDS\,)\cite{olmedo05}
(\url{http://spaceweather.gmu.edu/seeds/}), and the {Solar Feature Monitor}
\cite{qu06} (\url{http://filament.njit.edu/detection/vso.html}).

In this work, a multiscale edge detector is used to objectively identify and
track CME leading edges. In Section~3, {\it Transition Region and Coronal
Explorer} (\,TRACE; \opencite{handy99}\,) and {Large Angle and Spectrometric
COronagraph experiment} (\,SOHO/LASCO;  \opencite{brueckner95}\,) observations and
initial data-reduction is discussed, while the multiscale-based edge detection
techniques are presented in Section~4. Our results and conclusions are then
given in Sections~5 and 6.

\section{Observations} 
      \label{S-observations}      

To demonstrate the use of multiscale edge detection on CMEs in EUV imaging and
white-light coronagraph, data two data sets were used. The first data set
contains a CME observed on 18 April 2000 with the LASCO C2 and C3 telescopes.
The data set contains six C2 images and two C3 images. The second data set of
TRACE and LASCO observations contains a CME observed on 21 April 2002. The data
set contains one C2 image, three C3 images, and 30 TRACE images. TRACE
observed a very faint loop-like system propagating away from the initial flare
brightening, which was similar in shape to a CME (or number of CMEs) observed in
LASCO.  The appearance of the features remained relatively constant as they
passed through the TRACE 195~\AA\ passband and LASCO fields of view.

The TRACE observations were taken during a standard 195~\AA\ bandpass observing
campaign that provides 20 second cadence and an image scale of 0.5 arcsec per
pixel. Following \inlinecite{gallagher02b}, standard image corrections were first
applied before pointing offsets were accounted for by cross-correlating and
shifting each frame. The white-light LASCO images were obtained using a standard
LASCO observing sequence. The C2 images were taken using the orange filter (5400\,--\,6400~\AA) with
a variable cadence between 16 and 37 minutes and an image scale of 11.9 arcsec
per pixel. The C3 images were taken using the clear filter (4000\,--\,8500~\AA) with a variable
cadence between 24 and 60 minutes and an image scale of 56 arcsec per pixel.
Both the C2 and C3 images were unpolarized. Table 1 summarizes the details of these two data sets.
\begin{table}
\caption{Data sets for the 18 April 2000 and 21 April 2002 CMEs used in this work.}
\label{T-observations}
\begin{tabular}{|l| |l| |l| |l| |l| |l|}
\hline
 &\multicolumn{2}{l|}{18 April 2000}&\multicolumn{3}{l|}{21 April 2002}\\
\cline{2-6}
 &C2&{C3}&C2&{C3}&TRACE\\
\hline
Start time (UT) &16:06&17:18&01:27&01:42 &00:46 \\
Wavelength (\AA) &5400\,--\,6400&4000\,--\,8500&5400\,--\,6400&4000\,--\,8500&195\\
no. images & 6 & 2 & 1 & 3 &30\\
pix. size (arcsec) & 11.9 & 56 & 11.9 & 56 & 0.5\\
 min/max cadence & 16/37 min & 24/60 min & 16/37 min& 24/60 min & 20 sec\\
\hline
\end{tabular}
\end{table}

\section{Methodology}
  \label{S-method}
\subsection{Edge Detection}  
  \label{S-edgedetection}
Sharp variations or discontinuities often carry the most important information
in signals and images. In images, these take the form of boundaries described by
edges that can be detected by taking first and second derivatives \cite{marr82}.
The most common choice of first derivative for images is the gradient
\cite{gonzalez02}. The gradient of an image $I(x,y)$ at a point $(x,y)$ is the
vector,
\begin{equation}
  \label{grad}
  \grad{I}(x,y) = 
                    \left( \begin{array}{c} 
                    \frac{\partial{I}}{\partial{x}}\\ 
                    \frac{\partial{I}}{\partial{y}} 
                    \end{array} \right)
                 = 
                    \left( \begin{array}{c} 
                    G_{x}\\ 
                    G_{y} 
                    \end{array} \right).
\end{equation}
The gradient points in the direction of maximum change of $I$ at $(x,y)$. The
magnitude of this change is defined by the magnitude of the gradient, 
\begin{equation}
\label{gmag}
  |\grad{I}(x,y)| = (G_{x}^{2} + G_{y}^{2})^{1/2}. 
\end{equation}
The direction of the change at $(x,y)$ measured with respect to the $x$-axis is,
\begin{equation}
\label{angle}
\alpha(x,y) = \arctan(G_{x}/G_{y}). 
\end{equation}
The edge direction is perpendicular to the gradient's direction at
$(x,y)$.

The partial derivatives $G_x$ and $G_y$ are well approximated by the Sobel and
Roberts gradient operators \cite{gonzalez02}, although these operators cannot
easily be adapted to multiscale applications. In this  work, scale is considered
to be the size of the neighborhood over which the changes are calculated.

\begin{figure}    
  \centerline{\hspace*{0.015\textwidth}
               \includegraphics[width=1.0\textwidth,clip=]{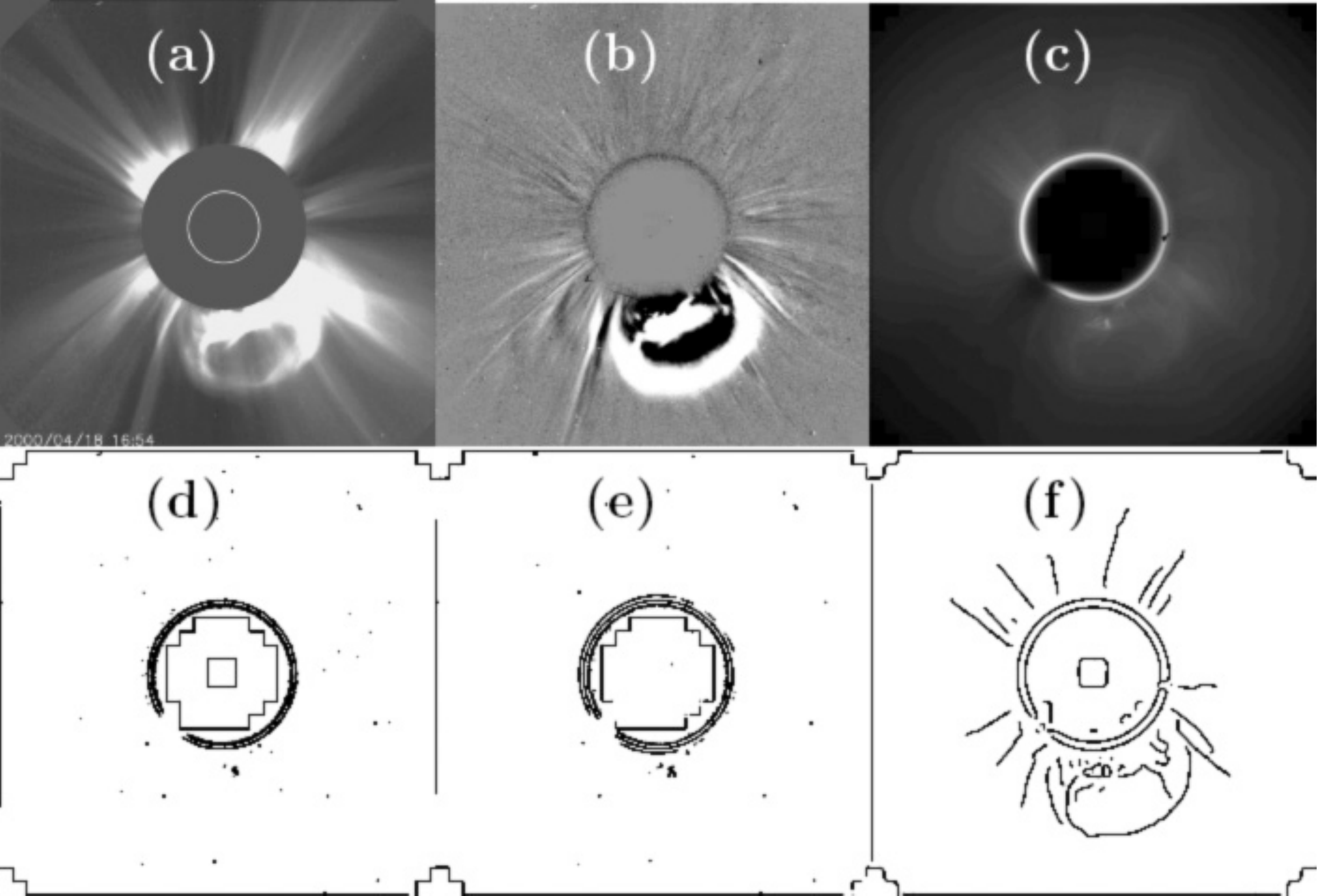}
               \hspace*{-0.03\textwidth}
                }
            
\caption{(a) A LASCO C2 image of a 18 April 2000 CME at 16:54 UT processed with
a monthly background model. (b) A running-difference image made by subtracting the
previous C2 image from the 16:54 UT image. (c) The unprocessed C2 image. (d)
Edges from applying the Roberts edge detector to the unprocessed image. (e)
Edges from applying the Sobel edge detector to the unprocessed image. (f) Edges
from applying the Canny edge detector ($\sigma = 5$) to the unprocessed image.
        }
   \label{F-firstfig}
   \end{figure}

Most edges are not steep, so additional information to that returned by the
gradient is needed to accurately describe edge properties. This can be achieved
using multiscale techniques. First proposed by \inlinecite{canny86}, this form of
edge detection uses Gaussians of different width ($\sigma$) as a smoothing
operator $\theta_{\sigma}$. The Gaussians are convolved with the original
image, so that Gaussians with smaller width correspond to smaller length-scales.
Equation (1) for the Canny edge detector can be written as,
\begin{equation}
  \label{grad}
  \nabla{(\theta_{\sigma}{\ast}I(x,y))} = 
                    \left( \begin{array}{c} 
                    \frac{\partial{}}{\partial{x}}{(\theta_{\sigma}{\ast}I(x,y))}\\ 
                    \frac{\partial{}}{\partial{y}}{(\theta_{\sigma}{\ast}I(x,y))} 
                    \end{array} \right).
\end{equation}
The image is smoothed using a Gaussian filter with a selected ${\sigma}$. Then a
derivative of the smoothed image is computed for the $x$ and $y$-directions. The
local gradient magnitude and direction are computed at each point in the image.
An edge point is defined as a point whose gradient-magnitude is locally maximum
in the direction defined by ${\alpha}$. These edge points form ridges in the
gradient magnitude image. The process of non-maximal suppression is performed by
setting to zero all pixels not lying along the ridges. The ridge pixels are then
thresholded using two thresholds ($T_a$ and $T_b$ with $T_b > T_a$). Ridge pixels
with values between $T_a$ and $T_b$ are defined as weak edges. The ridge pixels
with values greater than $T_b$ are called strong edges. The edges are linked by
incorporating the weak edges that are 8-connected with the strong pixels
\cite{gonzalez02}. Figure 1 shows a comparison of the Roberts, Sobel, and Canny
edge detectors applied to an unprocessed LASCO C2 image of the 18 April 2000 CME
from 16:54 UT. The Roberts (Figure 1(d)) and the Sobel (Figure
1(e)) detectors pick up a small piece of the CME core and noise. Using the
multiscale nature of the Canny detector (Figure 1(f)), choosing a larger scale size ($\sigma =
5$), edges corresponding to streamers and the CME front can be seen. 
Unfortunately, the Canny method has two main limitations: {\it{i}}) it is slow because
it is based on a continuous transform and, {\it{ii}}) there is no natural way to
select appropriate scales.

\subsection{Multiscale Edge Detection} 
      \label{S-msedgedetection}      

\inlinecite{mallat92a}, showed that the maximum modulus of the continuous wavelet transform
(\,MMWT\,) is equivalent to the multiscale Canny edge detector described in the
previous section. The wavelet transform converts a  2D image into a 3D function, where two of the
dimensions are position parameters and the third dimension is scale. The
transform decomposes the image into translated and dilated (scaled) versions of
a basic function called a wavelet: $\psi(x,y)$. A wavelet dilated by a scale factor $s$ is denoted as
\begin{equation}
 \psi_{s}(x,y) =  \frac{1}{s^2}\psi(\frac{x}{s},\frac{y}{s}) =  {s^{-2}}\psi({s^{-1}}{x},{s^{-1}}{y}).
\end{equation}

The wavelet is a function that satisfies a
specific set of conditions, but these functions have the key characteristic that
they are localized in position and scale \cite{mallat98}. The
minimum and maximum scales are determined by the wavelet transform, addressing the first problem of the 
Canny edge detector mentioned in the previous section. 

\inlinecite{mallat92b} refined the wavelet transform describe by
Mallat and Hwang, creating a fast, discrete transform. This addresses the computational speed
problem of the Canny edge detector making this fast transform more suited to realtime applications. As with the Canny edge
detector, this wavelet starts with a smoothing function: ${s^{-2}}\theta({s^{-1}}{x},{s^{-1}}{y})$. The
smoothing function is a cubic spline, a discrete approximation of a Gaussian. The smoothing function is also separable,
{\it i.e.} $\theta(x,y) = \theta(x)\theta(y)$. The wavelets are then
the first derivative of the smoothing function. This allows the wavelets to be
written as
\begin{equation}
  \psi_{s}^{x}(x,y) = {s^{-2}}\frac{\partial\theta({s^{-1}}x)}{\partial{x}}{\theta({s^{-1}}y)} \mbox{ and }  \psi_{s}^{y}(x,y) = {{s^{-2}}\theta({s^{-1}}x)}\frac{\partial\theta({s^{-1}}y)}{\partial{y}}.
\end{equation}
Another factor that adds to the speed of the wavelet transform algorithm is the
choice of a dyadic scale factor ($s$). Dyadic means that $s=2^j$ where
$j=1,2,{\ldots},J$ or $s=2^1, 2^2, 2^4, 2^8,{\ldots},2^J$, smallest scale to
largest scale. The index $J$ is determined by the largest dimension of the
image, $N$, {\it i.e.}, $N=2^J$. The wavelet transforms of $I(x,y)$ with respect to
$x$ and $y$ at scale {\it s} can then be written
\begin{equation}
\begin{array}{c}
W_{s}^{x}I = W_{s}^{x}I(x,y) = \psi_{s}^{x}(x,y)\ast I(x,y)\\
\mbox{and}\\
W_{s}^{y}I = W_{s}^{y}I(x,y) = \psi_{s}^{y}(x,y)\ast I(x,y),
\end{array}
\end{equation}
where $\ast$ denotes a convolution. Substituting these into Equations
(2) and (3) gives the following expression for the gradient of an image at scale {\it s} in terms
of the wavelets:
\begin{equation}
\label{wmag}
\begin{array}{c}
  |\nabla_{s}{I}(x,y)| = ((W_{s}^{x}I)^{2} + (W_{s}^{y}I)^{2})^{1/2}\\
  \mbox{and}\\
   \alpha_{s}(x,y) = \arctan(W_{s}^{x}I/W_{s}^{y}I).
\end{array}
\end{equation}
The detailed steps associated with implementing Equation\,(6) are shown in
Figure~2. The rows from top to bottom are scales one to five, respectively. Column (a)
displays the horizontal wavelet components $W_{s}^{x}I(x,y)$. Column (b) shows the
vertical wavelet components $W_{s}^{y}I(x,y)$. The next two columns show the magnitude
(c) of the multiscale gradient and the angle (d) of the multiscale gradient. The
edges calculated from the multiscale gradient are displayed in column (e).
\begin{figure}    
  \centerline{\hspace*{0.015\textwidth}
                \includegraphics[width=1.0\textwidth,clip=]{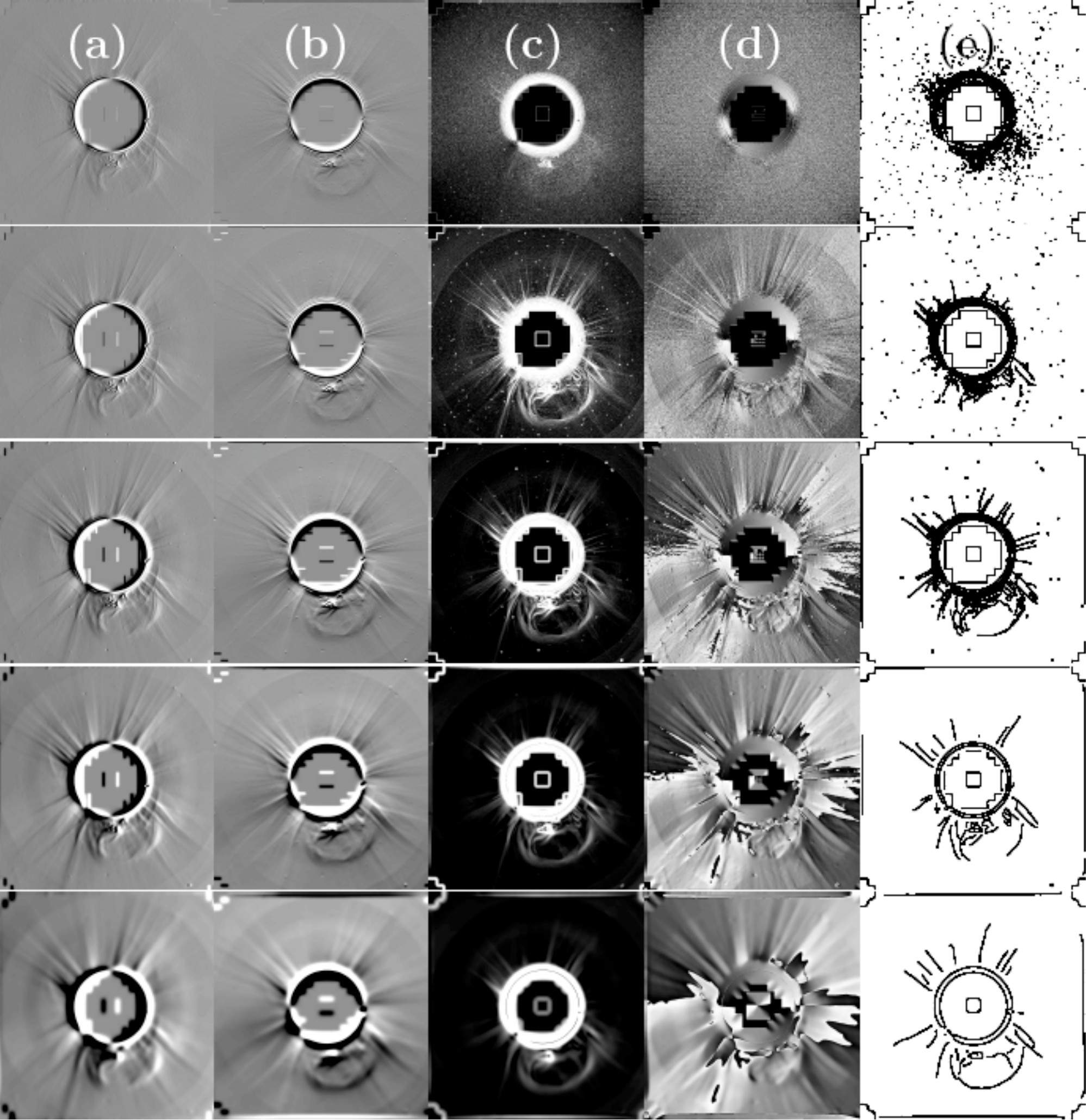}
              \hspace*{-0.03\textwidth}
                }
            
\caption{The steps calculating the multiscale edges of the
unprocessed C2 image (18 April 2000 CME at 16:54 UT). The rows from top to
bottom are scales $j=1$ to $j=5$ respectively. (a) Horizontal wavelet coefficients. (b)
Vertical wavelet coefficients. (c) The magnitude of the multiscale gradient. (d)
The angle of the multiscale gradient. (e) The edges calculated from the
multiscale gradient.
        }
   \label{F-25panels}
   \end{figure}

\subsection{Edge Selection and Error Estimation}  
  \label{S-edgeselect}

Once the gradient was found using the wavelet transform, the edges were
calculated using the local maxima at each scale, as described in the end of
Section 4.1. Closed or stationary edges due to features such as coronal loops,
ribbons, and cosmic rays were then removed, thus leaving only moving, open edges
visible. Currently not all available information is used so there were still some open spurious edges that were removed manually. 
Finally, only the edges from expanding, moving features were left. 
It was these edges that were used to characterize the temporal evolution of the CME front.

The multiscale edge detector can objectively define the CME front but it is also
important to estimate the statistical uncertainty in the edges and to obtain
errors in position or height. A straightforward way to do this is by using a
bootstrap (\inlinecite{efron93}). To do this we must create statistical realizations
of the data, then apply the multiscale edge detection method to each
realization. The realizations of the data are created by estimating a true,
non-noisy image then applying a noise model to the true image. Applying the
noise model means using a random number generator (random deviate) for our particular
noise model to generate noise and adding to the non-noisy image
estimate. In our case the noise model is well approximated by Gaussian noise.
Estimation of the noise and the true image is described by
\inlinecite{starck02}. The noise model is applied 1000 times to the true image with
each application creating a new realization. Doing this 1000 times, a mean and
standard deviation is calculated for the edge location or in our case for each
height point used. The steps for creating the estimate are {\it{i}}) estimate the
noise in the original image, {\it{ii}}) compute the isotropic wavelet transform of the
image ({\it \`a trous} wavelet transform), {\it{iii}}) threshold the wavelet
coefficients based on the estimated noise, and {\it{iv}}) reconstruct the image. The
reconstructed image is the estimate of the true image. The noise estimate from
the original image is used in the noise model applied to the estimated true
image.

\section{Results} 
  \label{S-results}

 \begin{figure}    
  \centerline{\hspace*{0.015\textwidth}
              \includegraphics[width=1.0\textwidth,clip=]{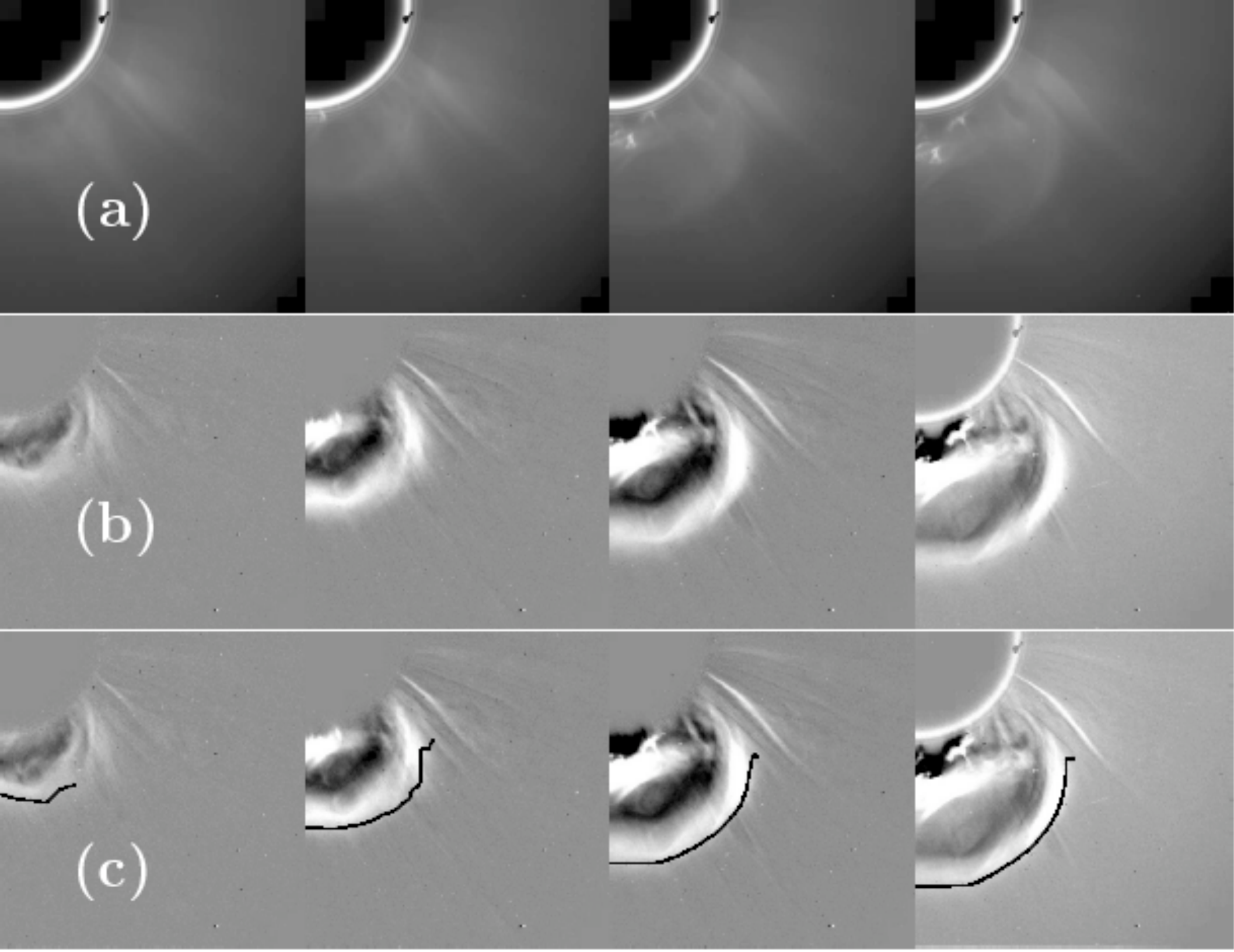}
              \hspace*{-0.03\textwidth}
                }
             
\caption{Illustration of a CME edge detection in subsequent images. (a) The
original LASCO C2 images, (b) running difference images of the LASCO C2 images, and (c)
application of the multiscale edge detection algorithm to the sequence of the
original images. The edges are the black lines displayed over the running difference
images. The CME erupted on 18 April 2000, the times for the frames are (from
left to right) 16:06 UT, 16:30 UT, 16:54 UT, and 17:06 UT. }
   \label{F-fig3}
   \end{figure}

The first example of application of the multiscale edge detection is illustrated
in Figure~3. Figure 3(a) shows four of the original, unprocessed LASCO C2 images
for the 18 April 2000 data set. The times for the frames from left to right are
16:06 UT, 16:30 UT, 16:54 UT, and 17:06 UT. Figure~3(b) shows running difference images
for the sequence of C2 images. The results of the multiscale edge detection
method applied to the original images are shown in Figure~3(c). The edges of the
CME front are displayed as black contours over the difference images shown in
Figure~3(b). Once the method is applied to the entire data set of C2 and C3
images, one point from each edge (all at the same polar angle) is selected. The
distance of each point from Sun center is plotted against time to create a
height-time profile. A bootstrap (described in Section ~4.3) was performed for
each edge calculation so that each point in the height--time profile has an
associated error in height.

 \begin{figure}    
  \centerline{
               \includegraphics[width=0.7\textwidth,angle=90,clip=]{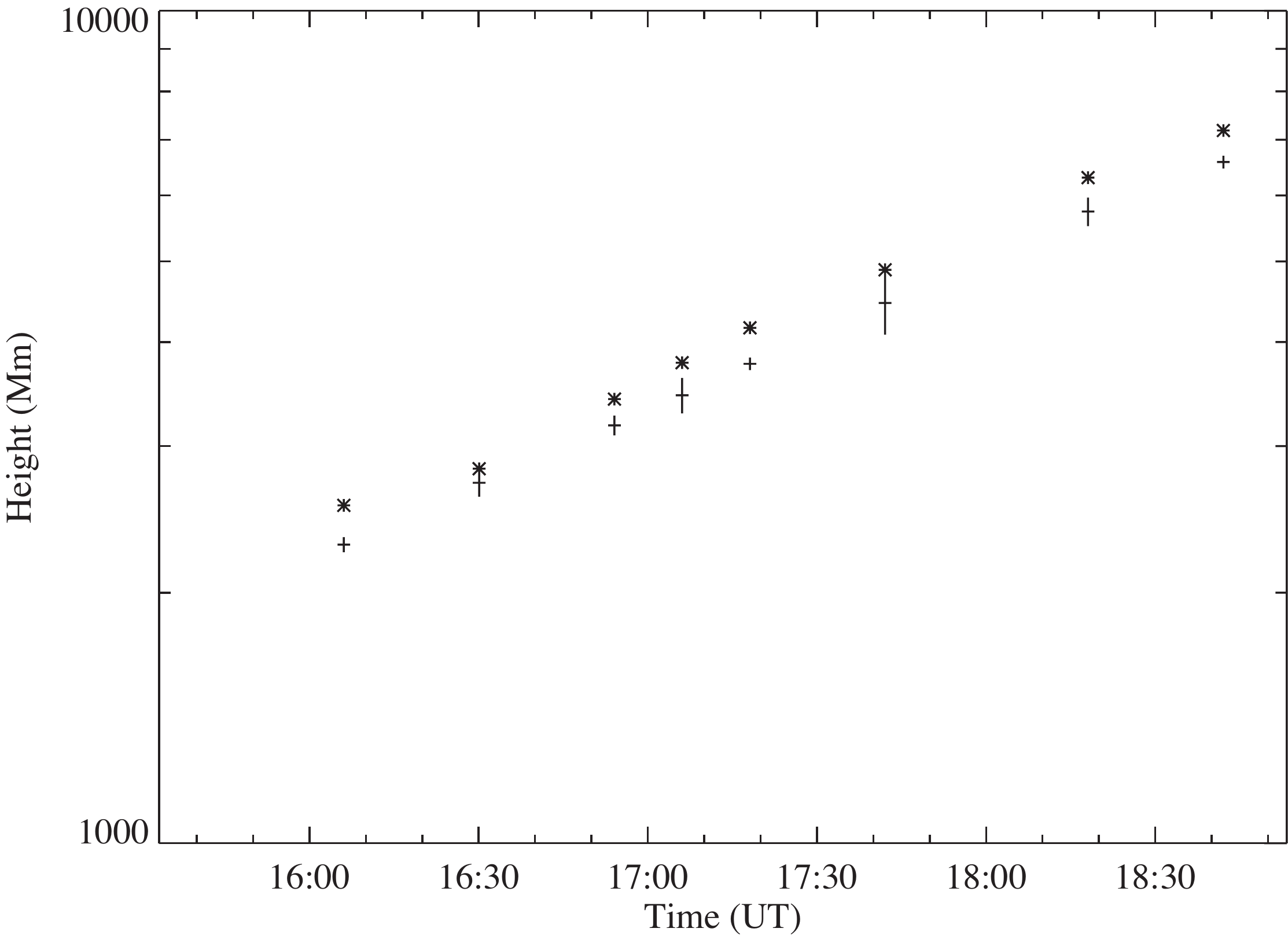}
                }
\caption{The height--time profile for the 18 April 2000 CME. The data plotted
using + symbols were obtained via the multiscale methods described herein,
while the data plotted using $\ast$ symbols are from the CU CDAW CME catalog. The
first six points are for LASCO C2 and the last two are for LASCO C3.}
   \label{F-fig4}
   \end{figure}

 The resulting height--time profile is shown in Figure~4. The data plotted using
 + symbols are those obtained via the multiscale method. One-${\sigma}$ errors in
 height are also shown. For comparison, data from the CUA CDAW CME catalog are
 plotted in the figure using $\ast$ symbols. The points determined with the
 multiscale method are systematically lower in height than the points from the
 CUA CDAW points. This is because the CUA CDAW points are selected using
 difference images and, as can be seen in Figure~3(c), the true CME front (black
 edge) is inside of the edge in the difference images. This illustrates a
 drawback of using difference images to determine the CME front. The difference
 image height estimates from the CUA CDAW catalog are larger than the multiscale edge estimates
 by $\approx{10}\%$.

\begin{figure}    
  \centerline{\hspace*{0.015\textwidth}
              \includegraphics[width=1.0\textwidth,clip=]{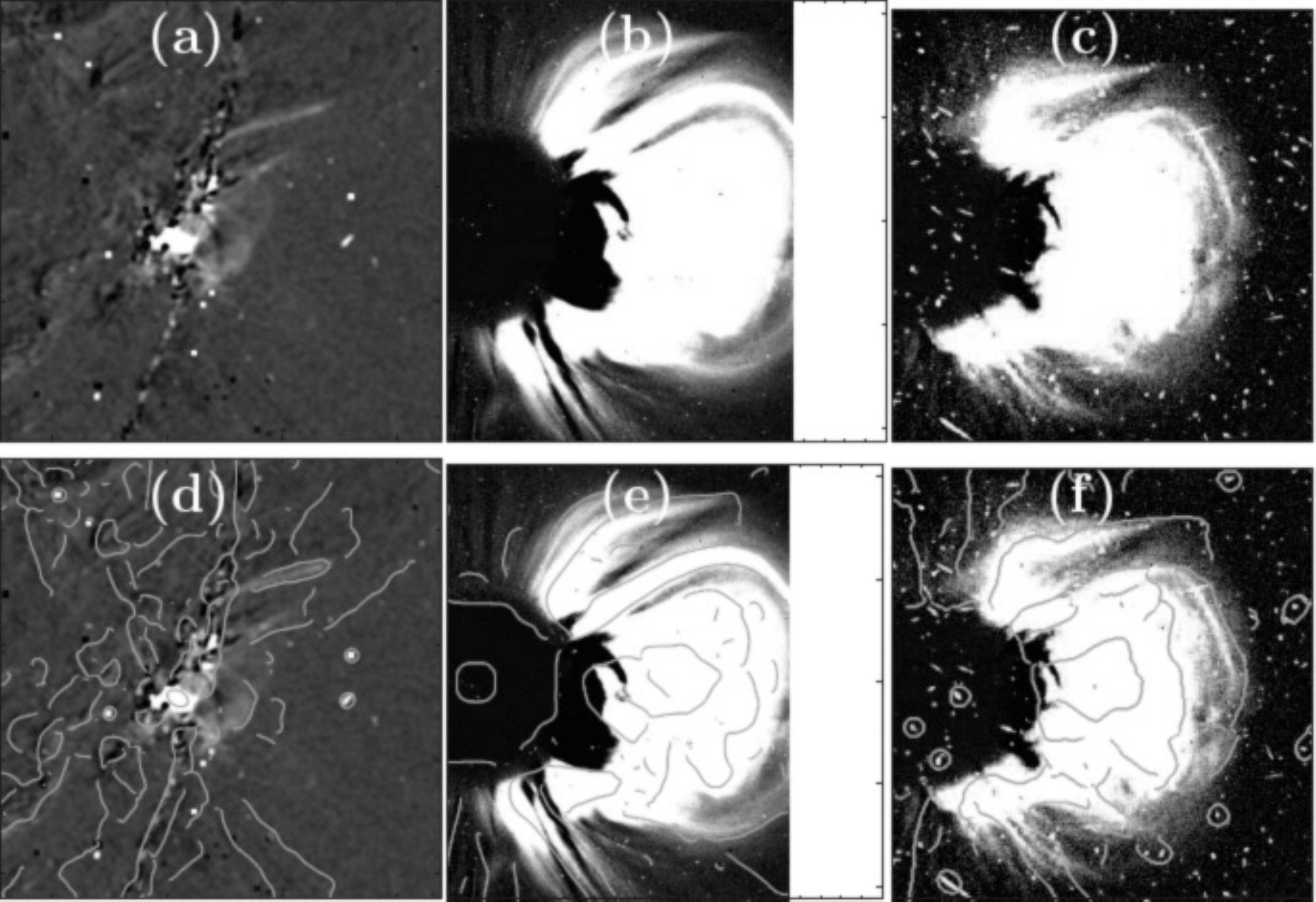}
              \hspace*{-0.03\textwidth}
                }

\caption{(a) TRACE 195~\AA\ difference image from 21 April 2002, created by
           subtracting an image taken at 00:43:10~UT from an image at
           00:42:30~UT, together with running difference images from (b) LASCO C2 (01:50:05~UT) and (c) LASCO C3 (02:13:05~UT).
           The same (d) TRACE and (e,f) LASCO images (as shown in (a), (b), and (c) respectively,
            but now overlaid with multiscale edges from a scale ($j=8$) that isolates the leading-edge of the CME. (The C2 images are not cut off on the right side.
            In order for the C2 images to have a similar FOV as the C3 images and for all images to be square, white space was added to the C2 images.)}
   \label{F-fig5}
   \end{figure}

Figure 5 displays data from the 21 April 2002 data set. Figure 5(a) shows a TRACE difference image created by subtracting an image
at 00:42:30~UT from an image at 00:43:10~UT. A very faint loop-like feature was only
visible in the difference image after it was smoothed and scaled to an extremely
narrow range of intensities. Both these operations were arbitrarily decided
upon, and are therefore likely to lead to the object's true morphology being
distorted. Figure 5(d) shows the same TRACE difference image as Figure 5(a) but overlaid with multiscale edges at scale $j = 8$. The edge of the faint
loop-like features is clearly visible as a result of decomposing the image using
wavelets, and then searching for edges in the resulting scale maps using the
gradient-based techniques described in the previous section. Figures 5(b)
and 5(c) show LASCO C2 and C3 running difference images, respectively. Figures 5(e)
and 5(f) show the LASCO difference images overlaid with the multiscale edges for
scale $j = 8$.

 \begin{figure}    
  \centerline{\hspace*{0.015\textwidth}
               \includegraphics[width=1.0\textwidth,clip=]{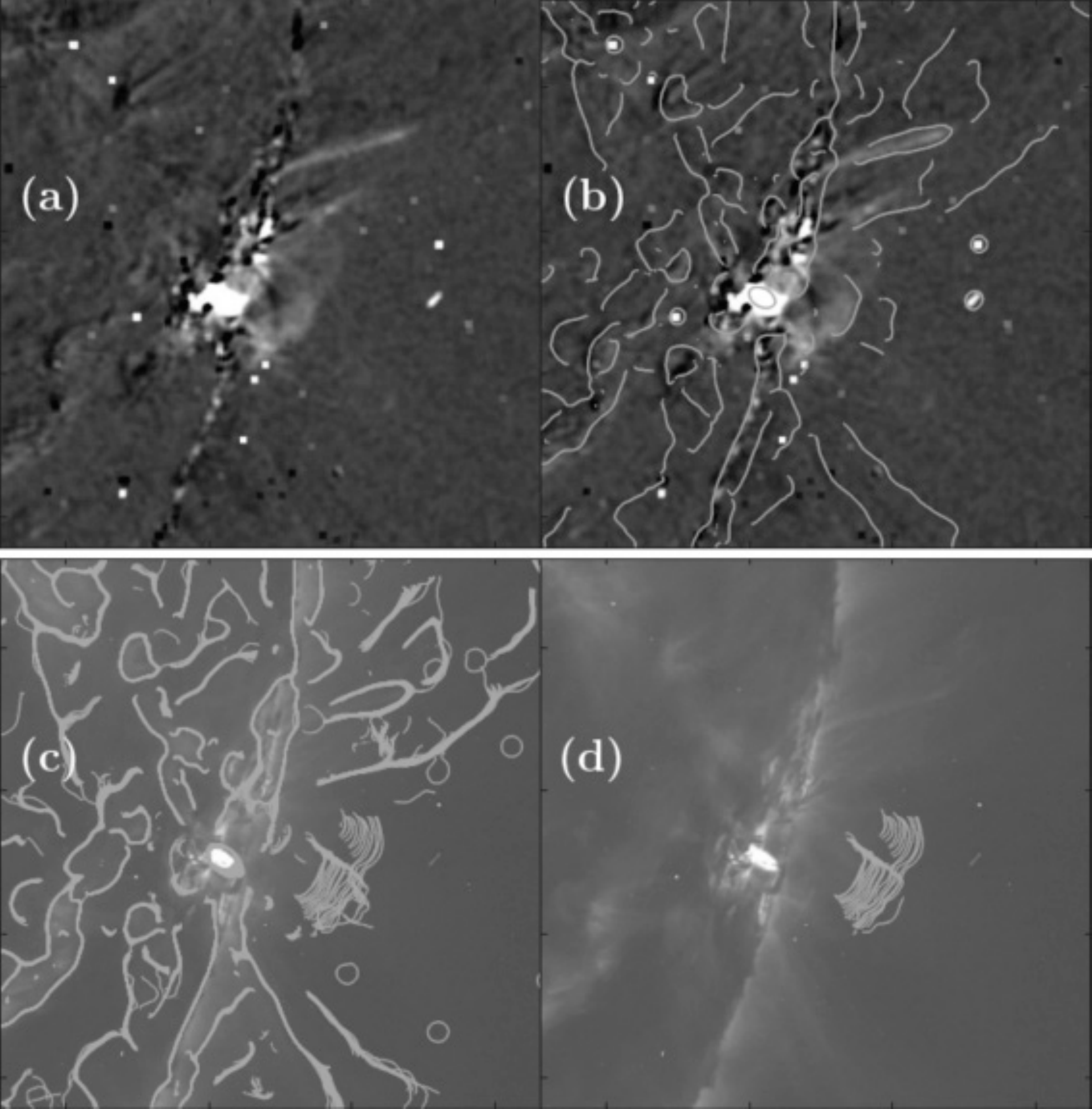}
              \hspace*{-0.03\textwidth}
                }
              
\caption{(a) TRACE 195~\AA\ difference image from 21 April 2002, created by
           subtracting an image taken at 00:43:10~UT from an image at
           00:42:30~UT. (b) The same TRACE images,
            but now overlaid with multiscale edges from a scale ($j = 8$) that isolates the
leading-edge of the CME.
            (c) The set of multiscale edges at scale $j = 8$, from all 30 TRACE images (00:46:34 UT to 01:05:19 UT)
             superimposed on the first image of the TRACE sequence. 
            (d) Same as Figure 6(c), but with unwanted edges removed. }
   \label{F-fig6}
   \end{figure}

Figure 6 displays on the TRACE data for the 21 April 2002 data set. Figure 6(a) displays the 
processed difference image (same as Figure 5(a)) and Figure 6(b) is the difference image
overlaid with multiscale edges at scale $j=8$ (same as Figure 5(d)). The underlying TRACE image in
Figure 6(c) is the original image from 00:46:34 UT. The multiscale edges at scale $j=8$ were
calculated for all 30 TRACE images from 00:46:34 UT to
01:05:19~UT. All 30 sets of edges were overlaid upon the original base image. Figure 6(d)
contains the same set of edges but by using size and shape information, the
expanding front is isolated. The leading edge, only partially visible in the
original TRACE difference image in Figure 6(a), is now clearly visible and
therefore more straightforward to characterize in terms of its morphology and
kinematics. The multiscale edges reveal the existence of two separate moving features.

\begin{figure}    
   \centerline{\includegraphics[width=1.0\textwidth,clip=]{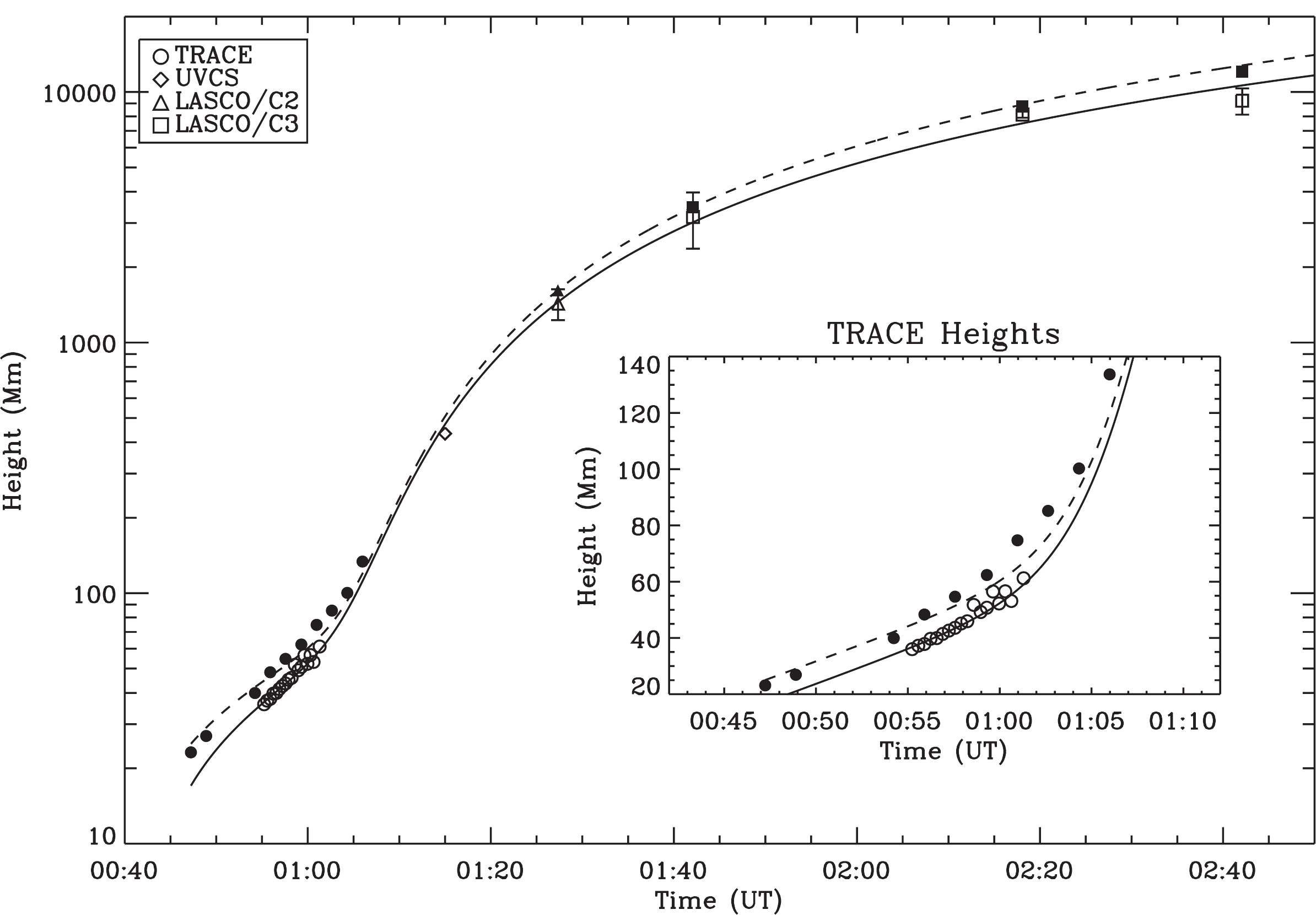}
                      }
\caption{The height--time profile for the 21 April 2002 CME, together with fits
assuming an exponential acceleration of the form given in Equation\,(9). The data
plotted using open symbols were obtained via the multiscale methods described
here, while the data plotted using solid symbols are
from Gallagher, Lawrence, and Dennis (2003). The solid line is a fit to the multiscale data, the
dashed to the latter. Error bars on the LASCO data have been multiplied by a
factor of two to improve their visibility and the mean uncertainty in
TRACE-derived heights are equal to the diameter of the open and filled circles
($\approx$3~Mm).}
   \label{F-fig7}
   \end{figure}

Using these edges, the expansion and motion of the CME from the sun is now
clearly visible and therefore characterizing it in terms of its morphology and
kinematics is more straightforward. The resulting height--time plot is show in
Figure~7, together with data from \inlinecite{gallagher03}, for comparison. We again find the heights determined manually are
larger by $\approx{10} \%$ using LASCO data and $\approx{20}
\%$ using TRACE data.

Following a procedure similar to that of Gallagher, Lawrence, and Dennis, the height-time curve
was fitted assuming a double exponential acceleration profile of the form:
\begin{equation}
a(t) = \left(\frac{1}{a_r \exp{(t/\tau_r)}} +
\frac{1}{a_d\exp{(-t/\tau_d)}}\right)^{-1},
\end{equation}
where $a_r$ and $a_d$ are the initial accelerations and $\tau_r$ and $\tau_d$
give the e-folding times for the rise and decay phases. A best fit to the
height--time curve was obtained with $h_0=17\pm3$~Mm, $v_0=40\pm4$~km~s$^{-1}$,
$a_r=1\pm1$~m~s$^{-2}$, $\tau_r=138\pm26 $~s, $a_d=4950\pm926$~m~s$^{-2}$, and
$\tau_d=1100\pm122$~s, and is shown in Figure~7.

\section{Conclusions and Future Work} 
 \label{S-conclusions}  
CMEs are diffuse, ill-defined features that propagate through the solar corona and
inner heliosphere at large velocities $(\geq$100~km~s$^{-1})$ \cite{yashiro04}, 
making their detection and characterization a difficult
task. Multiscale methods offer a powerful technique by which CMEs can be
automatically identified and characterized in terms of their shape, size,
velocity, and acceleration.

Here, the entire leading edge of the 18 April 2000 and 21 April 2002 CMEs has
been objectively identified and tracked using a combination of wavelet and
gradient based techniques. We have shown that multiscale edge detection successfully
locates the front edge for both well-defined events seen in LASCO as well as very faint structures
seen in TRACE. Although height--time profiles were only calculated for
one point, this method allows us to objectively calculate height--time profiles
for the entire edge. This represents an advancement over previous
point-and-click or difference-based methods, which only facilitate the CME apex
to be tracked. Comparing height--time profiles determined using standard methods
with the multiscale method shows that for these two CMEs the heights determined manually are
larger by $\approx{10} \%$ using LASCO data and $\approx{20}
\%$ using TRACE data.
 
Future work is needed to fully test the use of this technique in an automated system. Application of 
this edge detection method to a large, diverse set of events is necessary. An important improvement would 
be better edge selection. This will be accomplished by better
incorporating scale information. By chaining the edges together as a function of
scale we can distinguish false edges from true edges. This information can also
be used to better distinguish better different-shaped edges. Another improvement can be
made by using image enhancement. During the denoising stage, wavelet-based image
enhancement (such as in \opencite{stenborg03}) can be performed at the same time
that noise in the images is estimated and reduced. More sophisticated multiscale
methods using transforms such as curvelets will be studied. In order to
distinguish between edges such as those due to streamers from those due to a CME
front angle information can be incorporated. This multiscale edge detection has
been shown to have potential as a useful tool for studying the structure and dynamics of
CMEs. With a few more improvements this method could prove to be an important
part of an automated system.

\begin{acks}
The authors thank the referee for their suggestions and comments and
NASA for its support. CAY is supported by the NASA SESDA contract. PTG is supported by a grant from Science Foundation Ireland's Research Frontiers Programme.
\end{acks}


\end{article} 
\end{document}